\documentclass[%
reprint,
superscriptaddress,
 amsmath,amssymb,
 aps,
]{revtex4-2}

\usepackage{rotating}
\usepackage{overpic}
\usepackage{graphicx}
\usepackage{dcolumn}
\usepackage{bm}
\usepackage[utf8]{inputenc}
\usepackage[T1]{fontenc}
\usepackage{mathptmx}
\usepackage{etoolbox}

\usepackage{ragged2e}

\usepackage{float}
\usepackage[caption=false]{subfig}
\usepackage{dcolumn}

\newcommand{\mr}{\mathbf{r}}
\newcommand{\mft}{\tilde{\mathbf{f}}}

\newcommand{\tlo}{\tilde{\omega}}
\newcommand{\ud}{\mathrm{d}}

\begin{document}

\title{Influence of random surface deformations on the resonance frequencies and quality factors of optical cavities and plasmonic nanoparticles}

\author{Philip Trøst Kristensen}
\affiliation{Department of Electrical and Photonics Engineering, Technical University of Denmark, Ørsteds Plads, Building 343, 2800 Kgs. Lyngby, Denmark}%
\email{ptkr@dtu.dk}
\author{Thomas Kiel}
\affiliation{Institut f{\"u}r Physik, Humboldt-Universität zu Berlin, Newtonstraße 15, 12489 Berlin, Germany.}
\author{Kurt Busch}
\affiliation{Institut f{\"u}r Physik, Humboldt-Universität zu Berlin, Newtonstraße 15, 12489 Berlin, Germany.}
\affiliation{Max-Born-Institut, Max-Born-Straße 2a, 12489 Berlin, Germany}
\author{Francesco Intravaia}
\affiliation{Institut f{\"u}r Physik, Humboldt-Universität zu Berlin, Newtonstraße 15, 12489 Berlin, Germany.}

\begin{abstract}
\vspace{1mm}
Surface deformations of optical cavities and plasmonic nanoparticles are inevitable in nanophotonics. The random morphology changes of different realizations modify the associated resonance frequencies and quality factors, which may be characterized by specified 
distributions instead of their nominal values. 
%
As an alternative to statistical analyses based on direct numerical calculations, we present an approximate method using first-order perturbation theory with shifting boundaries. For an example resonator in the form of a plasmonic nanowire, the approach explains the bivariate frequency distribution observed in direct numerical calculations involving 1000 realizations of random surface deformations and provides the average and the associated covariance matrix with relatively high accuracy.
\end{abstract}


\maketitle
Nanophotonic devices based on optical cavities~\cite{Vahala_Nature_424_839_2003} or plasmonic resonators~\cite{Novotny_NP_5_83_2011} are technologically interesting for applications ranging from classical signal processing and sensing to quantum optical communication and computing. The designs often feature critical dimension of tens of nanometers or shorter, and fabrication limitations may therefore in practice lead to fabricated structures with an appreciable degree of surface deformations. The impact of such deformations -- which may be in the form of shifted resonance frequencies or added scattering losses -- must necessarily be assessed by comparing to the properties of the smooth structures: optical resonators may support resonances with narrow line shapes, which make them sensitive to deformations on a smaller relative length scale than plasmonic resonators, which typically feature much broader resonances. As examples, the impact of surface deformations on the finesse and quality factor of optical cavity arrays was investigated in Ref.~\cite{Dolan_OL_35_3556_2010}, and the influence of surface roughness on the optical properties of plasmonic nanoparticles was studied in Refs.~\cite{Rodriguez-Fernandez_PCCP_11_5909_2009,Trugler_PRB_83_081412_2011}. %
In addition to localized resonators, surface roughness has been studied in the context of planar optical waveguides~\cite{Marcuse_BellSys_48_3187_1969,Lacey_IEE_Proc_137_282_1990} and 
slow-light photonic crystal waveguides~\cite{Patterson_PRL_102_253903_2009}. Although surface deformations are often detrimental to the proper working of a given device, we note that they have also been exploited to enhance light trapping in thin-film photovoltaics~\cite{Hammerschmidt_Proc_SPIE_8620_2013,Zeman_SolEner_119_94_2013}. 
\begin{figure}[b!]
\vspace{-2mm}
\flushright
\begin{overpic}[width=.95\columnwidth]{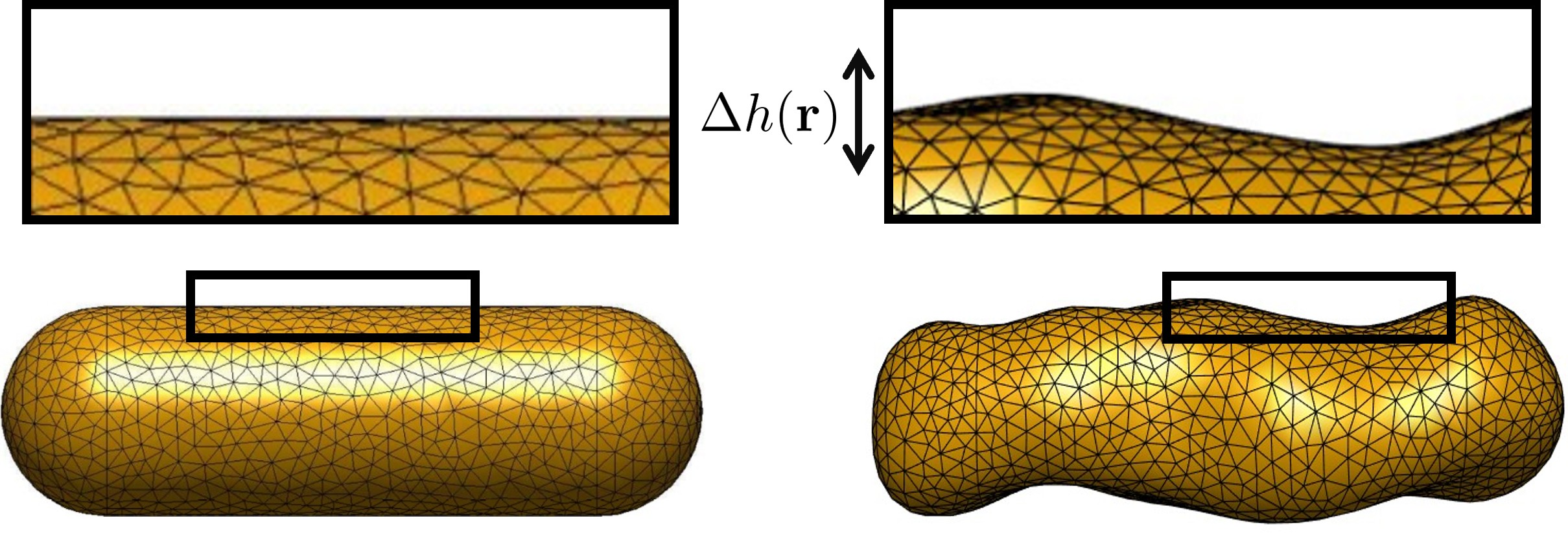}
\end{overpic}\\[2mm]
\begin{overpic}[width=.995\columnwidth]{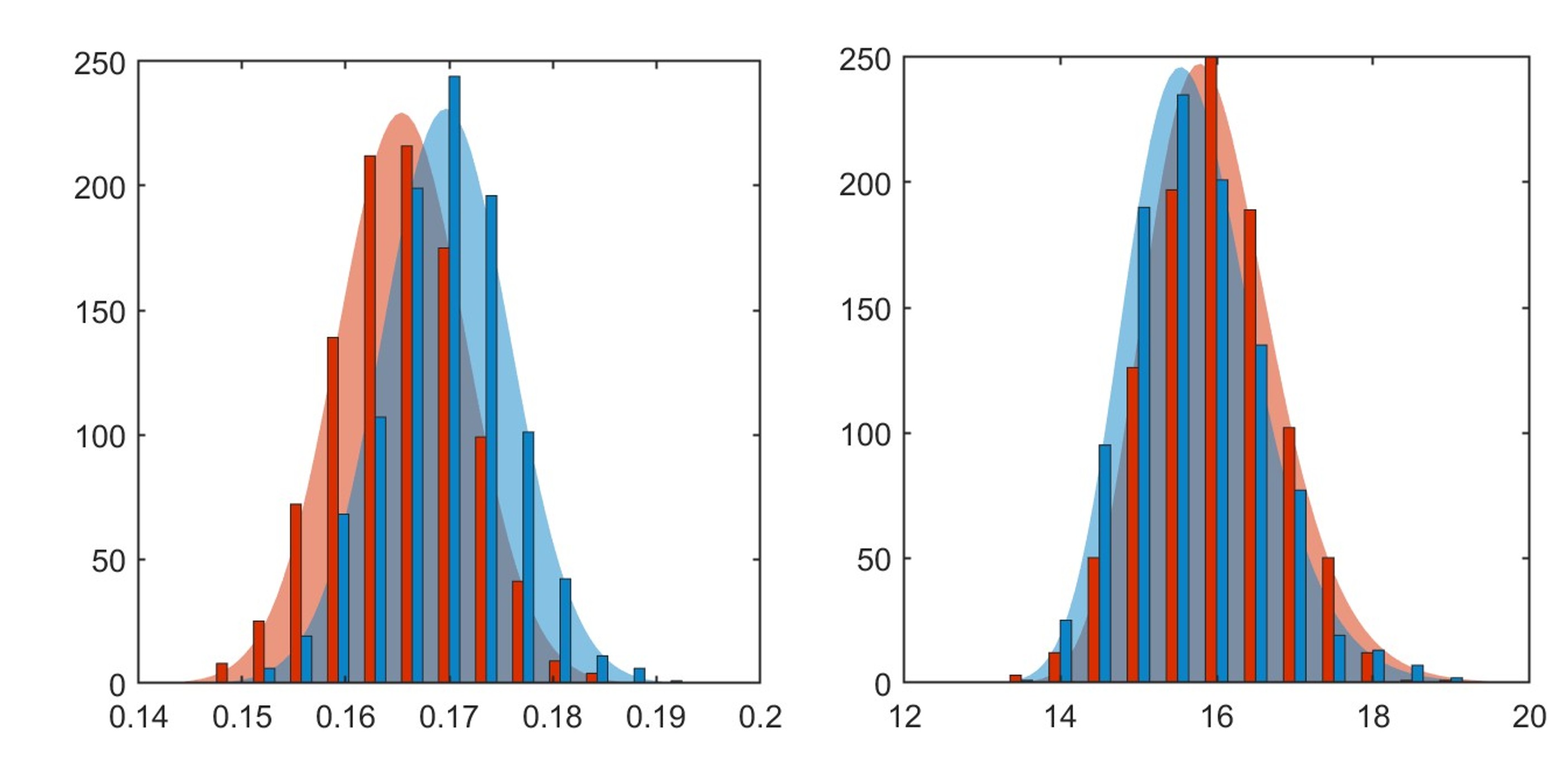}
\put(-1,19.5){\begin{sideways}Counts\end{sideways}}
\put(5,-4){Resonance frequency $\omega_\text{R} L/2\pi\text{c}$}
\put(66,-4){Quality factor $Q$}
\end{overpic}\\[3mm]
\caption{\label{Fig:fig4}\justifying Top: local surface deformations of a smooth nanowire (left) can be modeled as random local displacements $\Delta h(\mr)$ (right). Bottom: Histograms showing the resonance frequencies (left) and quality factors (right) of 1000 realizations of the deformed nanowire. Red and blue datasets correspond, respectively, to full numerical reference calculations and the semi-analytical approximate method introduced in this work. The shaded areas indicate the associated distributions.} 
\end{figure}

In this Letter, we focus on the impact of surface deformations on the resonance frequencies and quality factors of general electromagnetic resonators, such as optical cavities and plasmonic nanoparticles. Whereas full numerical calculations of individual resonators can be performed with high accuracy, a thorough statistical analysis on a representative ensemble of resonators with surface deformations is typically prohibitively expensive. As an alternative, we present a semi-analytical approach to investigate the statistics of resonators with deformed surfaces based on analysis of the corresponding smooth resonators. We illustrate the method using the example of a plasmonic nanorod for which Fig.~\ref{Fig:fig4} shows the statistical distribution of resonance frequencies and quality factors, but we note that the general approach applies equally well to optical cavities. 

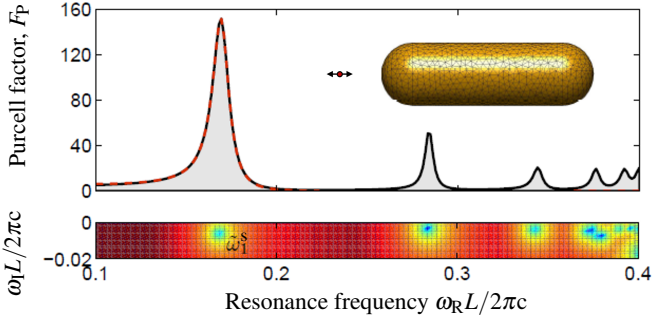
\begin{figure}[t!]
\flushright
\begin{overpic}[width=.95\columnwidth]{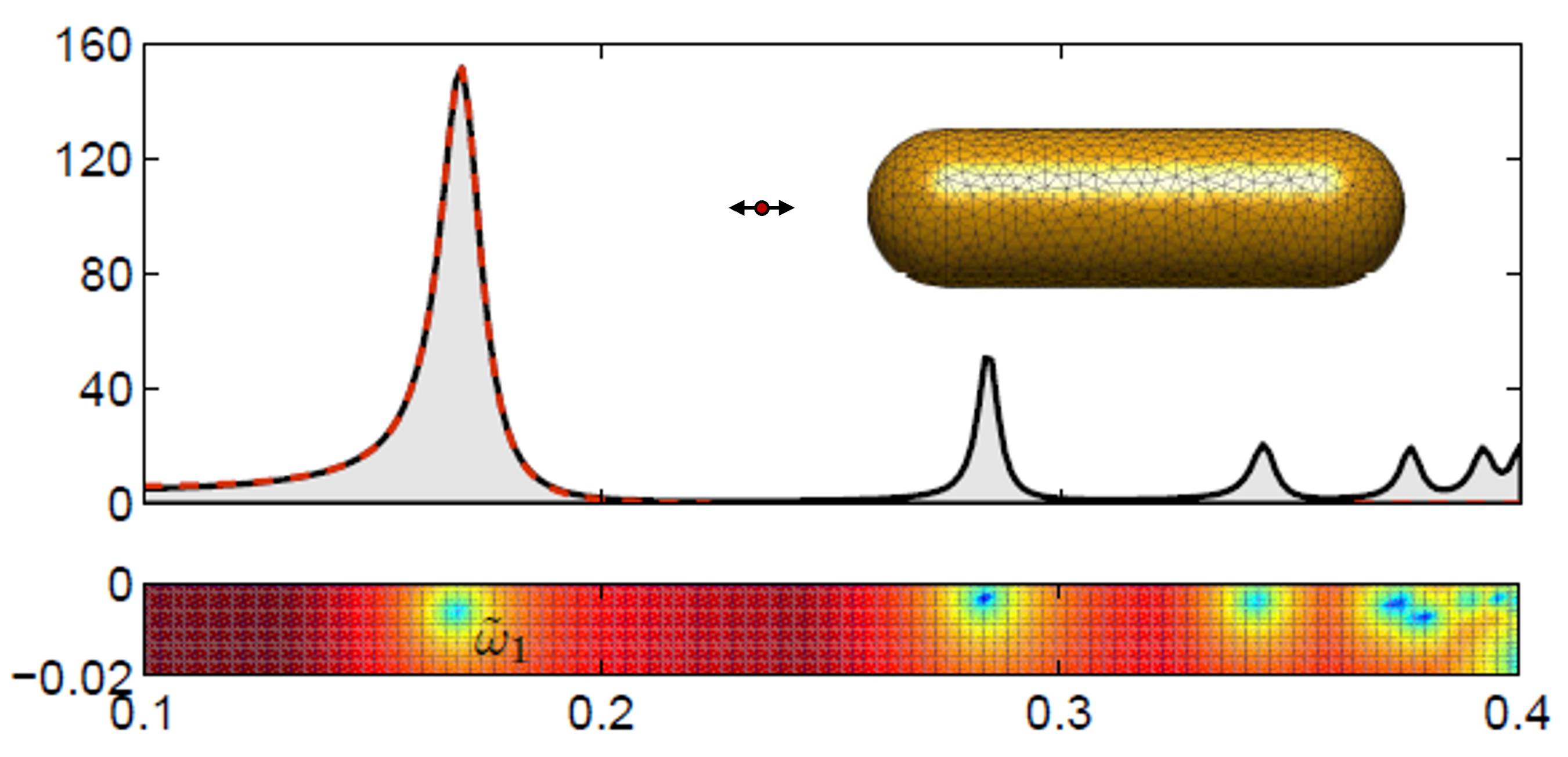}
\put(-5,20){\begin{sideways}Purcell factor, $F_\text{P}$\end{sideways}}
\put(32.5,8){\scriptsize \text{s}}
\put(30,-3){Resonance frequency $\omega_\text{R} L/2\pi\text{c}$}
\put(-6,-1){\begin{sideways}$\omega_\text{I}L/2\pi\text{c}$\end{sideways}}
\end{overpic}\\[1mm]
\caption{\label{Fig:Purcell_X_pos_m70_0_0_h_5_w_QNM}\justifying Top: Purcell factor for a point-like emitter 20 nm from the end of a smooth cylindrical gold nanowire (see inset) as a function of frequency. Red dashed curve shows the approximate response calculated using only the QNM  with complex frequency $\tlo_1^\text{s}$ as indicated below. Bottom: QNM spectrum in the plane of complex frequencies of the form $\omega_\text{R}+\text{i}\omega_\text{I}$, in which the discrete QNM resonance frequencies $\tlo_m$ show up as dark blue spots. } 
\vspace{-3mm}
\end{figure}

The dissipative modes of general electromagnetic resonators -- be they optical cavities or plasmonic nanoparticles -- can be accurately and precisely described by quasinormal modes (QNMs)~\cite{Ching_RevModPhys_70_1545_1998,Kristensen_ACS_Phot_1_2_2014,Lalanne_LPR_12_1700113_2018, Kristensen_AOP_12_612_2020}, also known in the literature as resonant states~\cite{Muljarov_EPL_92_50010_2010}. The QNMs are defined as the solutions to the wave equation subject to a suitable radiation condition to allow only fields traveling away from the resonator at large distances~\cite{Kristensen_AOP_12_612_2020}. The loss of energy in the form of radiation and possibly material absorption means that the QNMs have complex resonance frequencies of the form $\tlo_m=\omega_m-\text{i}\gamma_m$ with $\gamma_m>0$, from which the quality factor can be calculated directly as $Q_m= \omega_m/2\gamma_m$. For the analysis in this work, all QNM calculations were performed with the free boundary-element code MNPBEM~\cite{Hohenester_CPC_183_370_2012} and the iterative scheme of Ref.~\cite{Alpeggiani_SR_6_34772_2016}, see Ref.~\cite{Kristensen_AOP_12_612_2020} for details.  

As a generic example of an electromagnetic resonator of current interest in nanophotonics, we focus on a metallic nanowire with spherical end caps of total length $L=100\,$nm and radius $R=15\,$nm. We describe the metal by a Drude model of the form 
$\epsilon_\text{R}(\omega) = 1- \omega_\text{p}^2/(\omega^2+\text{i}\gamma\omega)$, %
where we use $\hbar\omega_\text{p}=7.9$eV and $\hbar\gamma=0.06$eV. Such nanowires, or variations thereof, are being investigated for use as nanoscale antennas for emission from colloidal quantum dots~\cite{Peter_ApplPhysB_124_83_2018} and were recently used in experiments demonstrating room-temperature strong-coupling to two-dimensional materials~\cite{Wen_NL_17_4689_2017}. The top panel of Fig.~\ref{Fig:Purcell_X_pos_m70_0_0_h_5_w_QNM} shows the Purcell factor of a point-like emitter at a position 20 nm away from the end of the nanowire in a configuration similar to that of Ref.~\cite{Peter_ApplPhysB_124_83_2018} with dipole moment in the direction of the nanowire. The Purcell factor represents the relative radiative decay rate of an emitter compared to free space, and the spectrum in Fig.~\ref{Fig:Purcell_X_pos_m70_0_0_h_5_w_QNM} is dominated by a number of distinct peaks. The characteristic resonances 
can be associated with the different QNMs supported by the nanowire, and this connection is directly visible when the positions of the resonance peaks are compared to the complex QNM spectrum. 
The calculation method employed here identifies the QNM resonance frequencies 
as the complex zeros of a characteristic function defined as the smallest magnitude eigenvalue of the numerical system matrix [12, 15]. The magnitude of this function is mapped in the bottom panel of Fig.~\ref{Fig:Purcell_X_pos_m70_0_0_h_5_w_QNM}, where the QNM frequencies show up as dark spots~\cite{Alpeggiani_SR_6_34772_2016,Kristensen_AOP_12_612_2020}. 
We focus on the QNM with complex resonance frequency $\tlo_1^\text{s}/\omega_L=0.16945(5)-0.00547(2)\text{i}$, where $\omega_L = 2\pi\text{c}/L$, and we use the superscript ``s'' to indicate that it is the frequency of the smooth cylinder. This frequency is that of the dipolar mode, which can be efficiently excited from the far field and can also act as an antenna to efficiently radiate energy away. The dashed red curve in Fig.~\ref{Fig:Purcell_X_pos_m70_0_0_h_5_w_QNM} shows the single-QNM approximation to the Purcell factor at the point of interest~\cite{Kristensen_AOP_12_612_2020}. 

For light-emission and antenna applications, one will often be interested in coupling light at a specific frequency. As we shall see, a direct influence of random surface deformations is to shift the complex resonance frequencies of individual realizations away from $\tlo_1^\text{s}$. As a consequence, the QNM resonance frequencies become themselves random variables with specific distributions to be determined. 

We model the surface deformations using a normally distributed random displacement $\Delta h(\mr)$ of the local resonator surface with zero mean and standard deviation $r_\text{rms}$ given as the root mean square of the displacements. Moreover, we assume that the local surface deformations are correlated on a length-scale of $\sigma$. For the present rather general analysis, we assume the surface deformations to result from the combined impact of numerous independent and random manufacturing processes. Invoking the central limit theorem~\cite{VanKampen_1992}, we therefore take the correlation function to be Gaussian, but we note that other correlation functions may be appropriate depending on the exact application. With these considerations, we %
write the average displacement and the autocorrelation function, respectively, as
\begin{align}
\langle\Delta h(\mr)\rangle & = 0\label{Eq:mean_h}\\[1mm]
\langle \Delta h(\mr)\Delta h(\mr')\rangle & = r_\text{rms}^2\text{e}^{-|\mr-\mr'|/2\sigma^2}.\label{Eq:variance_h} %
\end{align}

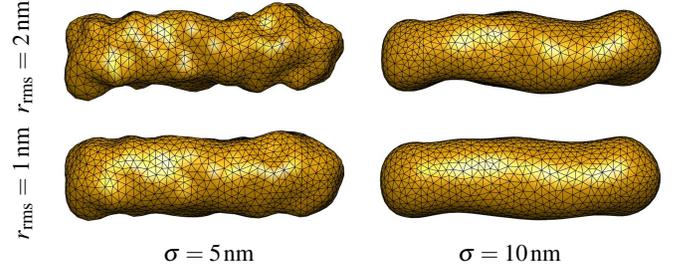
\begin{figure}[t!]
\flushright
\begin{overpic}[width=.925\columnwidth]{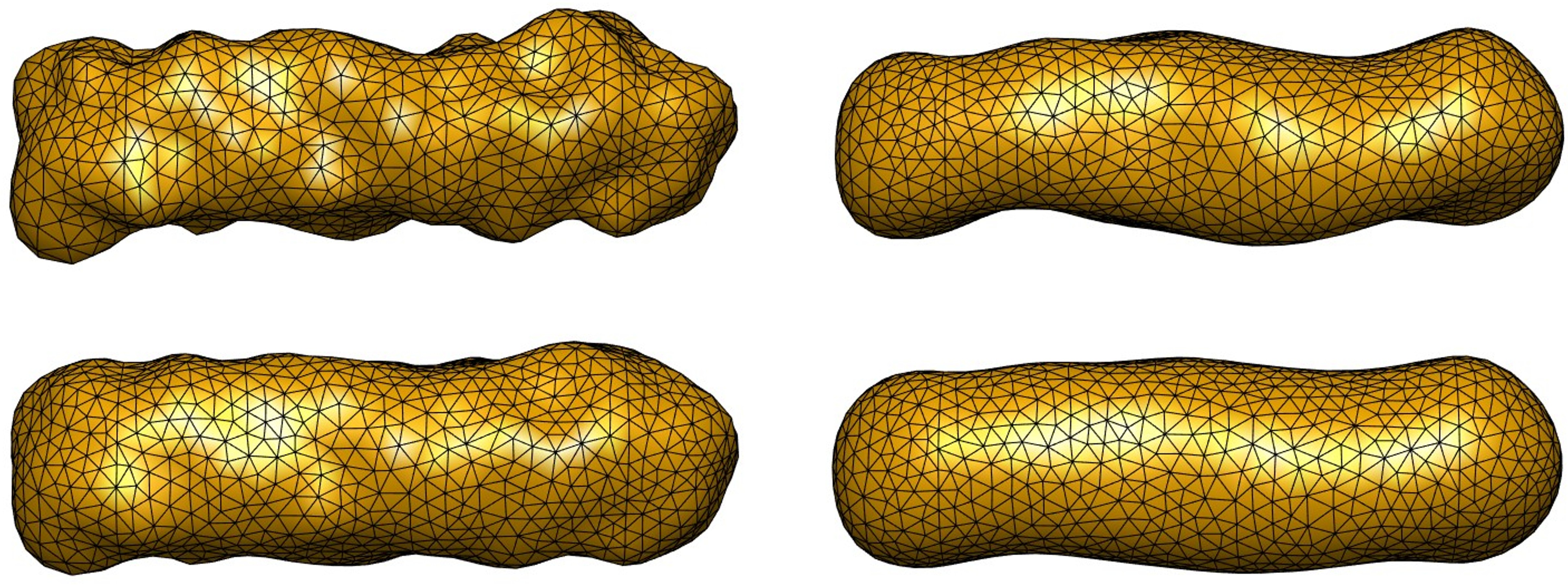}
\put(-8,18.5){\begin{sideways}$r_\text{rms}=2\,$nm\end{sideways}}
\put(-8,-2.5){\begin{sideways}$r_\text{rms}=1\,$nm\end{sideways}}
\put(17,-6){$\sigma=5\,$nm}
\put(66,-6){$\sigma=10\,$nm}
\end{overpic}\\[5mm]
\caption{\label{Fig:roughRoundedCylinders_4_diff_settings}\justifying Example meshes with random surface deformations drawn from distributions with standard deviation $r_\text{rms}$ and correlation length $\sigma$, as indicated. The examples were generated from the mesh of the smooth nanowire in the inset of Fig.~\ref{Fig:Purcell_X_pos_m70_0_0_h_5_w_QNM} by normally distributed random displacements with specified autocorrelation as defined in Eq.~(\ref{Eq:variance_h}).} 
\vspace{-2mm}
\end{figure}

For surface deformations in the direct numerical calculations,  
we use the approach of 
Ref.~\cite{Loth_JOSAB_40_B1_2023}. We first prepare a smooth mesh using the meshing tool ``Gmsh''~\cite{gmsh} and subsequently shift the individual vertices to create a random shape deformation with the specified characteristics. We have found that this approach works well as long as the correlation length is much larger than the mesh size, which in practice is always the case, since the mesh size also governs the numerical error, as discussed below. Figure~\ref{Fig:roughRoundedCylinders_4_diff_settings} shows examples of four different realizations of the surface deformations with $r_\text{rms}=1\,$nm or $r_\text{rms}=2\,$nm and $\sigma=5$nm or $\sigma=10\,$nm. The examples were calculated with the same seed for the random number generator, which results in identical distributions of positive and negative local deformations in the top and bottom row. The height of the local deformations in the top row is double that in the bottom row as it directly scales with $r_\text{rms}$. The average distance between local deformations is governed by $\sigma$, and therefore it is larger in the right column. 

\begin{figure}[t!]
\flushright
\begin{overpic}[width=.925\columnwidth]{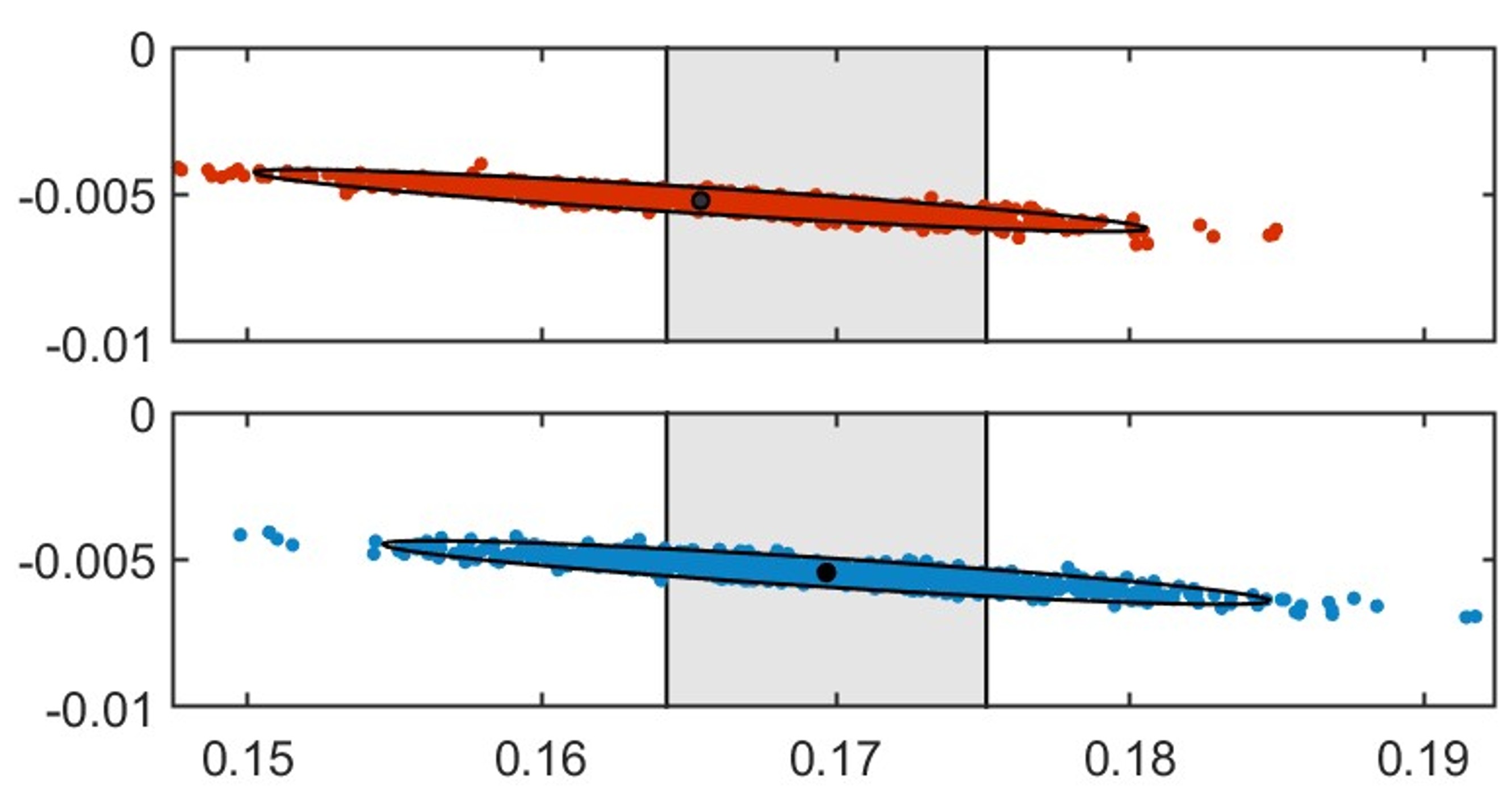}
\put(-5,21){\begin{sideways}$\omega_\text{I}L/2\pi\text{c}$\end{sideways}}
\put(29.5,-3.5){Resonance frequency $\omega_\text{R} L/2\pi\text{c}$}
\end{overpic}\;\\[2.5mm]
\caption{\label{Fig:3}\justifying Top: distribution of complex QNM resonance frequencies $\tlo_1$ from 1000 realizations of the surface deformations with $r_\text{rms}=2\,$nm and $\sigma=10\,$nm. The light gray shading indicates the full width at half max of the peak corresponding to $ \tlo_1^\text{s}$ in Fig~\ref{Fig:Purcell_X_pos_m70_0_0_h_5_w_QNM}. The small circle marks the mean $\tilde{\mu}_1=0.1654-0.0052\text{i}$, 
and the black ellipse marks the boundary of the 95\% confidence interval. Bottom: 1000 random samples drawn from a distribution with specified average and covariance matrix calculated by the approximate method.} 
\vspace{-2mm} 
\end{figure}

For the specific case of $r_\text{rms}=2\,$nm and $\sigma=10\,$nm, the top panel of Fig.~\ref{Fig:3} shows the distribution of $\tlo_1$ resulting from 1000 different realizations of the surface deformations. The points are plotted in the plane of complex frequencies of the form $\tlo_\text{R}+\text{i}\omega_\text{I}$ along with the 95\% confidence contour of a fitted bivariate probability density function of the form
\begin{align}
f(\tlo_m) = \frac{1}{2\pi\sqrt{\text{det}\{{\Sigma}_m\}}}\text{e}^{-\frac{1}{2}(\underline{x}_m-\underline{\mu}_m)^{\sf T}{\Sigma}_m^{-1}(\underline{x}_m-\underline{\mu}_m)},
\end{align}
where $\underline{x}_m=[\text{Re}\{\tlo_m\},\text{Im}\{\tlo_m\}]^{\sf T}$, and $\underline{\mu}_m$ and ${\Sigma}_m$ denote the mean 
and the covariance matrix, respectively. For this case, we find
\begin{align}
\hspace*{-1.5mm}
\underline{\mu}_1\!/\!\omega_L\!=\!
\begin{bmatrix}
\!0.1654\\[1mm] \!-0.0052
\end{bmatrix}\;
\text{and}\;\;{\Sigma}_1\!/\!\omega_L^2=10^{-5}\!\times\!
\begin{bmatrix}
\! 3.8549 &   \! \!-0.2423 \\[1mm]
\!-0.2423  &  \! \!0.0185
\end{bmatrix}\!\!.\!\!
\label{Eq:mu_and_sigma_1000_realizations}
\end{align}

Numerical calculation results are known to be influenced by discretization errors, which in this case will also shift the calculated resonance frequencies around in the complex plane. In order for the calculated results to be meaningful, one must therefore make sure that statistics of the shifts due to surface deformations are significant compared to the expected numerical error. Because of the flat facets of the triangular mesh and the pulse basis functions used for the discretization~\cite{Hohenester_CPC_183_370_2012}, the convergence in this case is expected to be linear in the characteristic triangle side length $l$. The stated numbers in Eq.~(\ref{Eq:mu_and_sigma_1000_realizations}) are the raw numerical data from the fitting of 1000 points, each with numerical uncertainty stemming from the calculations with $l=3$nm. To gauge the level of uncertainty, we can compare the calculated value using $l=3$nm for the smooth cylinder, $\tlo_1^{3\text{nm}}/\omega_L=0.1697 - 0.0054\text{i}$, to the best estimate of the true value as stated above, which was obtained from a convergence study with varying $l$~\cite{Kristensen_AOP_12_612_2020}. This difference is less than a few parts in ten thousand, which is hardly visible on the scale of Fig.~\ref{Fig:3}. Gray shading in the figure shows the full width at half max of the peak in the Purcell factor-spectrum in Fig~\ref{Fig:Purcell_X_pos_m70_0_0_h_5_w_QNM}. The given level of surface deformations is sufficient to generate a distributions of frequencies which is wider than the linewidth.

Given the distribution in the complex plane and the associated bivariate fit, we can 
plot the histograms and distributions of real resonance frequencies and quality factors as shown by the red datasets in Fig.~\ref{Fig:fig4}. We note that the data points corresponding to the quality factors result from ratios of two normally distributed and correlated random numbers and therefore in general follow a non-trivial distribution~\cite{Marsaglia_JSS_16_1_2006}. 

High accuracy QNM calculations are computationally expensive, and it may not always be practical or feasible to carry out a full statistical analysis of the influence of surface deformations. With moderate computational efforts, however, one can estimate the influence of surface deformations using first-order perturbation 
theory for Maxwell's equations with shifting boundaries
~\cite{Lai_PRA_41_5187_1990,Johnson_PRE_65_066611_2002,Sztranyovszky_PRR_5_013209_2023}. This approach has been applied to estimate the change in resonance frequency of a plasmonic dimer of gold spheres~\cite{Kristensen_AOP_12_612_2020} and a photonic cavity~\cite{Kountouris_OE_30_40367_2022}, and we note that it was recently exploited to set up an elegant quality factor-optimization scheme in Ref.~\cite{Granchi_ACSphot_10_2808_2023}.

To set the stage for the statistical analysis, we write the complex frequency change $\Delta\tlo_m$ induced by the deformation using a two-element vector of real numbers as
\begin{align}
\underline{\Delta}_m
= \begin{bmatrix}
\text{Re}\{\Delta\tlo_m\}\\[1mm] \text{Im}\{\Delta\tlo_m\}
\end{bmatrix} 
= \int_{\partial V} \underline{F}_m(\mr)\Delta h(\mr)\ud A,
\label{Eq:Delta_m_integral}
\end{align}
in which $\partial V$ denotes the surface of the smooth nanowire. The vector function $\underline{F}_m(\mr) = [\text{Re}\{\tilde{F}_m(\mr)\},\text{Im}\{\tilde{F}_m(\mr)\}]^{\sf T}$ is defined by the weight function
~\cite{Lai_PRA_41_5187_1990,Johnson_PRE_65_066611_2002,Sztranyovszky_PRR_5_013209_2023}
\begin{align}
\tilde{F}_m(\mr) = &-\frac{\tlo_m}{2}\Big[\left(\epsilon_\text{R}(\omega)-1\right)\mft_m^{\|}(\mr)\cdot\mft_m^{\|}(\mr)\nonumber\\
&\qquad- \left(\epsilon_\text{R}^{-1}(\omega)-1\right)\mft_m^{\perp}(\mr)\cdot\mft_m^{\perp}(\mr)\Big],
\label{Eq:functionFm}
\end{align}
which contains the parallel ($\|$) and perpendicular ($\perp$) components of the $m$'th QNM field distribution $\mft_m(\mr)$ evaluated on the outside surface of the smooth nanowire. Combining Eqs.~(\ref{Eq:mean_h}) and (\ref{Eq:Delta_m_integral}), we find that the first moment of the complex frequency change vanishes, 
\begin{align}
\langle\underline{\Delta}_m\rangle = \int_{\partial V} \underline{F}_m(\mr)\langle\Delta h(\mr)\rangle\ud A = \underline{0}
\label{Eq:average_perturbative},
\end{align}
so that the approximate mean is equal to the complex QNM frequency of the smooth resonator. The covariance matrix is equal to the second moment, which we can express by Eqs.~(\ref{Eq:variance_h}) and (\ref{Eq:Delta_m_integral}) as 
\begin{align}
\langle\underline{\Delta}_m^2\rangle 
&= r_\text{rms}^2\int_{\partial V}\int_{\partial V'} \underline{F}_m(\mr)\underline{F}_m^{\sf T}(\mr')\text{e}^{-|\mr-\mr'|^2/2\sigma^2}\ud A\ud A'.
\label{Eq:correlation_perturbative}
\end{align}

To enable a direct comparison with the reference calculations, we now set $m=1$, $r_\text{rms}=2\,$nm and $\sigma=10\,$nm. From Eqs.~(\ref{Eq:average_perturbative}) and (\ref{Eq:correlation_perturbative}) we find the approximate mean and covariance matrix to be $\tilde{\mu}_1^\text{pert} = \tlo_1^\text{s} + \langle\Delta\tlo_1\rangle = \tlo_1^\text{s}$, 
and 
\begin{align}{\Sigma}_1^\text{pert}/\omega_\text{L}^2=\langle\underline{\Delta}_1^2\rangle/\omega_\text{L}^2 = 10^{-5}\times
\begin{bmatrix}
3.8037 &  -0.2422 \\[1mm]
-0.2422 &   0.0195
\end{bmatrix},
\label{Eq:Sigma_pert_best estimate}
\end{align}
respectively. The bottom panel of Fig.~\ref{Fig:3} shows a sample of random QNM resonance frequencies drawn from a distribution with specified mean and covariance given by $\tilde{\mu}_1^\text{pert}$ and $\boldsymbol{\Sigma}_1^\text{pert}$.  
%
%
%
%
Comparing to the reference calculations in the top panel, we see a relative shift  on the mean QNM frequency on the order of 2.5\%. This shift appears to depend strongly on the relative magnitude of the %
surface deformation characteristics. 
For the case of $r_\text{rms}=1\,$nm and $\sigma=5\,$nm we find a relative error on the order of 1.5\%, and choosing $r_\text{rms}=1\,$nm and $\sigma=10\,$nm, we find a relative error less than 1\% (not shown). Consequently, we attribute the missing shift to the approximate nature of the first-order perturbation theory. For the present case, the shift is well within the overall distribution of approximate frequencies, as seen by comparing the two panels of Fig.~\ref{Fig:3}. For other resonators, such as high-$Q$ optical cavities (not shown), the shift may be larger than the overall distribution and downwards in the complex plane to move the average quality factor to lower values as a result of additional scattering losses. Comparing Eqs.~(\ref{Eq:mu_and_sigma_1000_realizations}) and (\ref{Eq:Sigma_pert_best estimate}), we find that that the covariance matrix is reproduced with a similar level of accuracy as the mean. This observation is further corroborated by the direct comparison of the two ellipses in the panels of Fig.~\ref{Fig:3}.
%

%
%
%
%
%

%

\begin{figure}[t!]
\vspace{-2mm}
\flushright
\begin{overpic}[width=.975\columnwidth]{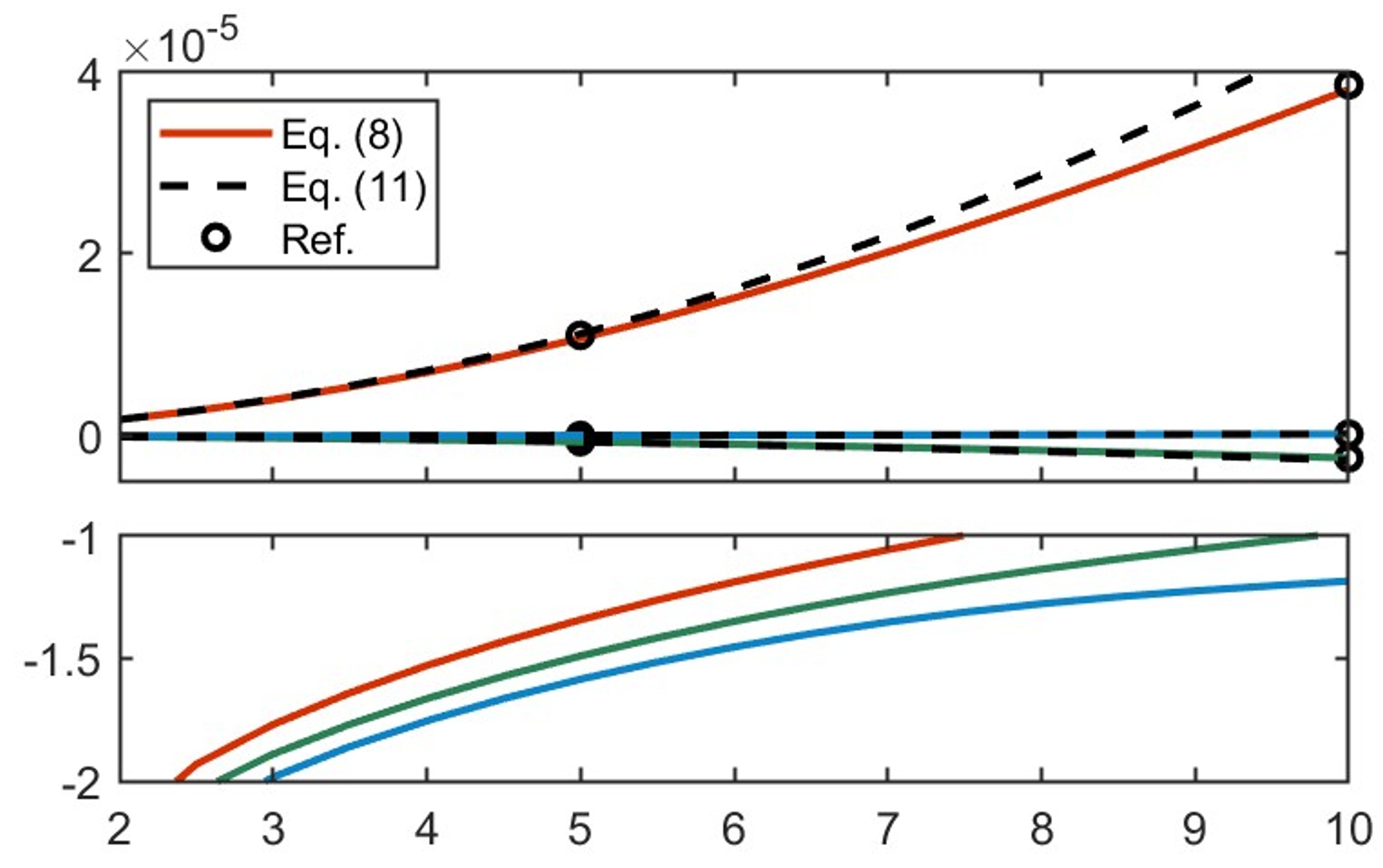}
\put(-3,39.5){\begin{sideways}${\Sigma}_1/\omega_L^2$\end{sideways}}
\put(-3,7){\begin{sideways}$\log_{10}\Delta{\Sigma}_\text{rel}$\end{sideways}}
\put(48,-3){$\sigma$ [nm]}
\end{overpic}\\[2mm]
\caption{\label{Fig:roundedCylinder_QNM_perturbative_smooth_varSigma}\justifying Top: Variation in the elements of the approximate covariance matrix as calculated by Eq.~(\ref{Eq:correlation_perturbative}) for fixed choice of $r_\text{rms}=2\,$nm and varying $\sigma$. Red, green, and blue lines indicate the matrix elements $\Sigma_{11}$, $\Sigma_{12}$, and $\Sigma_{22}$, respectively, dashed lines indicate the corresponding elements as calculated by Eq.~(\ref{Eq:correlation_perturbative_approx}), while markers show the results of full reference calculations with 1000 realizations. Bottom: Logarithm of the relative difference $\Delta{\Sigma}_\text{rel}$ between Eqs.~(\ref{Eq:correlation_perturbative}) and (\ref{Eq:correlation_perturbative_approx}).}\end{figure}

Based on the bivariate approximate distribution, we can again plot the resulting histograms and distributions of the approximate real resonance frequencies and quality factors as shown in the blue dataset in Fig.~\ref{Fig:fig4}. The small shift in the average resonance frequency is visible in the maxima of the distributions in the left panel. At the same time, it is much less pronounced 
in the distribution of quality factors as a consequence of the similarity in shape of the two distributions.
 
From Eq.~(\ref{Eq:correlation_perturbative}) we see that the covariance matrix is directly proportional to the square of the standard deviation, i.e., the variance, and to leading order it is also proportional to the square of the correlation length. 
%
To analyze this dependence, we consider the limit of small correlation lengths compared to the variations in the function $F_m(\mr)$, which is largely determined by the effective wavelength, cf. Eq.~(\ref{Eq:functionFm}). In these cases, we can approximate the Gaussian correlation function with the delta-function as
\begin{align}
\text{e}^{-|\mr-\mr'|^2/2\sigma^2}\approx2\pi\sigma^2\delta(|\mr-\mr'|),
\end{align}
and we note that similar approximations can be derived for other types of correlation functions~\cite{Lindell_Am_J_Phys_61_438_1993}. Inserting in Eq.~(\ref{Eq:correlation_perturbative}) we find then a much simpler expression for the approximate covariance matrix,
\begin{align}
{\Sigma}_1^\text{pert} &\approx 2\pi r_\text{rms}^2\sigma^2\int_{\partial V} \underline{F}_m(\mr)\underline{F}_m^{\sf T}(\mr)\ud A,
\label{Eq:correlation_perturbative_approx}
\end{align}
which notably involves just a single surface integral and therefore is comparable in complexity to the typical level of post-processing in numerical nanophotonics.

To illustrate the level of accuracy, the top panel of Fig.~\ref{Fig:roundedCylinder_QNM_perturbative_smooth_varSigma} shows a direct comparison of Eqs.~(\ref{Eq:correlation_perturbative}) and (\ref{Eq:correlation_perturbative_approx}) for varying correlation length. In the limit of small correlation length, the simple expression correctly reproduces the functional form of the perturbative analysis. We note also, that the differences between the approximate approach and the reference calculations are hardly visible on this scale. The bottom panel shows the relative error in using Eq.~(\ref{Eq:correlation_perturbative_approx}), compared to Eq.~(\ref{Eq:correlation_perturbative}), which scales as $\sigma^4$ consistent with the next term in the delta function-expansion of the symmetric gaussian correlation function~\cite{Lindell_Am_J_Phys_61_438_1993}.

As a result of the perturbative nature of the proposed approximate scheme, the accuracy is strongly connected with 
the magnitude of the local shape deformations. For the main calculations in this paper, we deliberately chose a relatively large value of the standard deviation of more than 10\% of the radius of the nanowire to investigate the limits of the proposed method. In many practical applications, we expect the surface deformations to be significantly smaller, in which case the resulting approximate resonance frequency distributions are expected to be correspondingly closer to the reference calculations, and in many cases of practical relevance, the simpler result in Eq.~(\ref{Eq:correlation_perturbative_approx}) may provide sufficient accuracy with practically no computational overhead. The fact that the suggested perturbative approach relies only on the QNM of the smooth structure means that the approach is useful also in cases where the practical calculation method does not easily allow one to perturb the mesh. In combination with traditional perturbation theory for estimates of the influence of the material model, for example, an analysis of the influence of surface deformations as presented in this Letter may find use as a general and relatively simple tool to assess the robustness of practical calculations when analyzing resonance frequencies and quality factors of electromagnetic resonators.






%
%

\bibliography{qnmbib.bib}  



\end{document}